\documentclass[10pt]{article}

\usepackage{graphicx}

\usepackage{cite}

\usepackage{color} 
\usepackage{amsmath,amssymb,graphics,graphicx} 

\usepackage[pdftex,bookmarks,colorlinks,breaklinks]{hyperref}  
\usepackage{setspace}
\usepackage{multirow}
\usepackage{color}
\usepackage{rotating}
\usepackage{verbatim}

\usepackage[labelfont=bf,labelsep=period,justification=raggedright]{caption}

\makeatletter
\renewcommand{\@biblabel}[1]{\quad#1.}
\makeatother

\date{}

\begin{document}
\begin{flushleft}
{\Large
\textbf{Kidney branching morphogenesis under the control of a ligand-receptor based Turing mechanism}
}
\vskip1cm
Denis Menshykau$^{1}$, Dagmar Iber$^{1,2\ast}$\\
1 Department for Biosystems Science and Engineering, ETH Zurich, Basel, Switzerland\\
2 Swiss Institute of Bioinformatics, Basel, Switzerland\\[2cm]

$^\ast$ Corresponding Author:\\
 Dagmar Iber\\
 Department for Biosystems Science and Engineering (D-BSSE)\\
 ETH Zurich \\
 Mattenstrasse 26\\
 4058 Basel\\
 Switzerland\\[0.5cm]
 +41 61 387 32 10 (phone)\\
+41 61 387 31 94 (fax)\\
 E-mail: dagmar.iber@bsse.ethz.ch \\[1.cm]

Accepted to: \emph{Physical Biology}
\end{flushleft}

\clearpage

\section*{Abstract}
The main signalling proteins that control early kidney branching have been defined. Yet the underlying mechanism is still elusive. We have previously shown that a Schnakenberg-type Turing mechanism can recapitulate the branching and protein expression patterns in wildtype and mutant lungs, but it is unclear whether this mechanism would extend to other branched organs that are regulated by other proteins. Here we show that the GDNF-RET regulatory interaction gives rise to a Schnakenberg-type Turing model that reproduces the observed budding of the ureteric bud from the Wolffian duct, its invasion into the mesenchyme, and the observed branching pattern. The model also recapitulates all relevant protein expression patterns in wild-type and mutant mice. The lung and kidney models are both based on a particular receptor- ligand interaction and require: (1) cooperative binding of ligand and receptor, (2) a lower diffusion coefficient for the receptor than for the ligand, and (3) an increase in the receptor concentration in response to receptor-ligand binding (by enhanced transcription, more recycling or similar). These conditions are met also by other receptor-ligand systems. We propose that ligand-receptor based Turing patterns represent a general mechanism to control branching morphogenesis and other developmental processes.

\emph{Keywords}: Computational Biology, Developmental Biology, Turing pattern, Branching Morphogenesis, Kidney Branching;

\clearpage


\section{Introduction}
Many organs, such as the lung, kidney, and glands are heavily branched structures.  At the organ level, the branching ÔÔpatternÕÕ is defined by several parameters, such as the site and type of branching (terminal versus lateral branching, bifurcation versus trifurcation), branch angles, rates of elongation, and changes in tubular diameter. All of these processes are controlled to yield an overall pattern unique to each organ. The branched tree in the lung is generated by the sequential, non-random use of three geometrically simple modes of branching (lateral branching, planar and orthogonal bifurcation), which occur in defined routines \cite{Metzger:2008ky}. Trifurcations have also been documented in the lung \cite{Blanc:2012ea}, but these are much more prevalent in the ureteric bud of the kidney.  Culture experiments revealed that most branching events in the kidney are terminal bifurcations and to a lesser extent trifurcations, and only $6\%$ of all branching events are lateral branching events  \cite{Watanabe:2004kr, Meyer:2004de, Costantini:2010p43730}.

It is a long-standing question how branching is controlled during development and whether the mechanism is the same in the different organs in spite of the differences in the architecture and in the regulatory proteins. We have recently proposed a model to explain the branching processes in the lung \cite{Menshykau:2012kg}. The model focused on the two key signalling factors that have been identified experimentally in the developing lung, fibroblast growth factor (FGF)10 and Sonic Hedgehog (SHH) as well as the SHH receptor PTCH1. We showed that the reported biochemical interactions give rise to Schnakenberg-type Turing patterns \cite{MurrayBook}  that result in a distribution of FGF10 as would be expected for lateral branching and bifurcations. The model also reproduced all published mutant phenotypes, including the counter-intuitive widening of the clefts between buds as \textit{Fgf10} expression is reduced in an allelic sequence \cite{Mailleux:2005jk,Ramasamy:2007hu}. In further simulations we showed that the same core regulatory network together with FGF9 is capable of controlling the emergence of smooth muscles in the clefts between growing lung buds, and \textit{Vegfa} expression in the distal sub-epithelial mesenchyme \cite{Celliere:2012jc}. In the model the experimentally observed upregulation of \textit{Fgf10} expression by FGF9 \cite{delMoral:2006p42857} promoted lateral branching over the bifurcation mode of branching. Regulatory networks involving FGF10 and SHH also control the branching in the prostate \cite{Wilhelm:2006p45512, Donjacour:2003ty}, the salivary gland \cite{Hsu:2010ep}, and in the pancreas \cite{Bhushan:2001ua, Pulkkinen:2003uy}. However, the regulatory proteins are different in the developing kidney \cite{Affolter:2009p25219,OchoaEspinosa:2012ux}, and the organs all differ in their branching pattern and geometry such that it is an open question whether the underlying regulatory principles are the same or different.

The ureteric bud forms at the beginning of embryonic day (E)10.5 at the level of the hindlimb in a specialized region called the metanephric mesenchyme \cite{Costantini:2010p43730}. The metanephric mesenchyme produces inductive signals that causes the nephric duct (also called Wolffian duct) to evaginate and to form a single ureteric bud near its caudal end \cite{Costantini:2010p43730}.  After invading the metanephric mesenchyme, the ureteric bud undergoes about ten generations of repeated branching \cite{Cebrian:2004ia}, followed by a period of elongation, then by one to two rounds of branching before birth \cite{Cebrian:2004ia}. Similar to the lung, the kidney collecting ducts form via branching and elongation of an epithelial cell layer \cite{Majumdar:2003tp}. Growth factors and components of the extracellular matrix are required, while the mesenchyme is dispensable \cite{Qiao:1999ui}. The mesenchyme, however, influences the pattern of branching. One striking example came from tissue recombination experiments in which lung mesenchyme induced branching of the ureteric bud with a pattern characteristic of lung epithelium, i.e. with increased lateral branching \cite{Lin:2001vf}. 

Based on these observations the local expression of growth promotors and inhibitors has long been considered to determine the branching pattern. In particular, given the chemoattractive properties of the  TFG-beta family protein Glial cell line-derived neurotrophic factor (GDNF) \cite{Tang:2002it,Tang:1998vf}, it has been suggested that branching of the ureteric bud is caused by the attraction of tips toward local sources of GDNF \cite{Sariola:2003jo}. GDNF signals to the cell via the receptors RET and GPI-anchored subunit GDNF family receptor alpha-1 (GFR$\alpha$1). Ureteric bud outgrowth fails in \emph{Gdnf}$^{-/-}$ and \emph{Ret}$^{-/-}$, \emph{Gfr}$\alpha^{-/-}$ mutant mice \cite{Treanor:1996dq, Costantini:2010p43730, Majumdar:2003tp, Pichel:1996en, Sanchez:1996cy}. Moreover, beads soaked with GDNF induce the outgrowth of extra ureteric buds in kidney culture explants \cite{Pepicelli:1997jz}. Conditional knock-outs further revealed a positive feedback between GDNF/RET and WNT11 signalling as part of the core branching mechanism  \cite{Majumdar:2003tp}. Expression of \textit{Wnt11} is activated in the epithelial tip of the ureteric bud and WNT11 signalling is in turn required to propagate mesenchymal GDNF signalling, which results in the establishment of an autoregulatory epithelial-mesenchymal feedback signalling loop \cite{Majumdar:2003tp}. FGF signalling, on the other hand, has only a supporting function in the ureteric bud.  Thus, branching is still observed in \emph{Fgfr2}-mutant kidneys, though at a reduced level \cite{SimsLucas:2009bq}. Moreover, enhanced FGF signalling in knock-out mice of the antagonist \emph{Sprouty} can rescue \emph{Gdnf}-/- and \emph{Ret}-/- mutants, which otherwise fail to develop kidneys \cite{Michos:2010p43732}. Interestingly, GDNF/RET and FGF10/FGFR have the same downstream target, the ETV4/ETV5 transcription factors \cite{Lu:2009p43731}, and it is therefore plausible that they may serve similar functions. 

The role of bone morphogenetic proteins (BMPs) in kidney development is less clear. Appreciable levels of phosphorylated SMADs are first detectable at E12.5 in collecting duct trunks and loss of SMAD1 transcriptional activation of WIF1 is associated with reduced \textit{Wif1} expression and increased WNT/$\beta$-catenin signalling activity in lung epithelia, resulting in specific fetal lung abnormalities \cite{Xu:2011ka}. Addition of BMPs enhances branching at low concentrations and inhibits branching at high concentrations \cite{Bush:2004uc, Clark:2001gp, Piscione:1997uba}. Equally, removal or misexpression of the BMP receptor \textit{Alk3} affects the branching pattern. Thus, expression of constitutively active \textit{Alk3} in collecting ducts using the \textit{HoxB7} enhancer-promoter reveals an inhibitory effect of BMP pathway activation on collecting ducts, confirming these \textit{in vitro} findings \cite{Hu:2003ti}. Moreover, removal of the BMP antagonist \textit{Gremlin (Grem)1} in mice results in excessive BMP activity in the metanephric mesenchyme around the ureteric bud and disrupts the invasion of the mutant metanephric mesenchyme by the ureteric bud and the concurrent establishment of the autoregulatory GDNF/WNT11 feedback signalling loop \cite{Michos:2007p42720}. In spite of these demonstrated effects of BMPs, removal of the common signal transducer  \textit{Smad4} within this tissue has little impact on kidney development and branching morphogenesis up to E16.5. The signals of the TGF$\beta$ superfamily must therefore be transduced through SMAD4- independent pathways in the collecting duct  \cite{Oxburgh:2004p42447}. To that end it was observed that BMP-2 antagonizes WNT signalling in osteoblast progenitors by promoting an interaction between SMAD1 and Dishevelled-1(DVL-1) that restricts $\beta$-catenin activation \cite{Liu:2006p5513}. 

Given its detailed characterization and obvious differences to the lung we have focused on branching morphogenesis in the ureteric bud to define common principles in branching morphogenesis. While the signalling circuits in the developing kidney are still incompletely characterized, the available data is sufficient to build a first computational model of the regulatory interactions. We show in the following that the reported biochemical interactions between GDNF and its receptor can give rise to a similar Schnakenberg-type Turing mechanism as observed for the SHH-PTCH1 interaction in the lung. We show that the model is consistent with all available wildtype and mutant data and that it reproduces the main modes of branching, i.e. bifurcations and trifurcations, that are observed during kidney branching morphogenesis. Finally we show that the same regulatory network is capable of controlling the epithelial invasion of the metanephric mesenchyme. We argue that the coupling of a receptor-ligand based Turing patterning mechanism with the impact of the domain geometry may constitute a general mechanism to control branching morphogenesis.

\section{Results}

\subsection{A Computational Model for the Regulatory Network}

\begin{figure}[h]
\begin{center}
\includegraphics[width=1\textwidth]{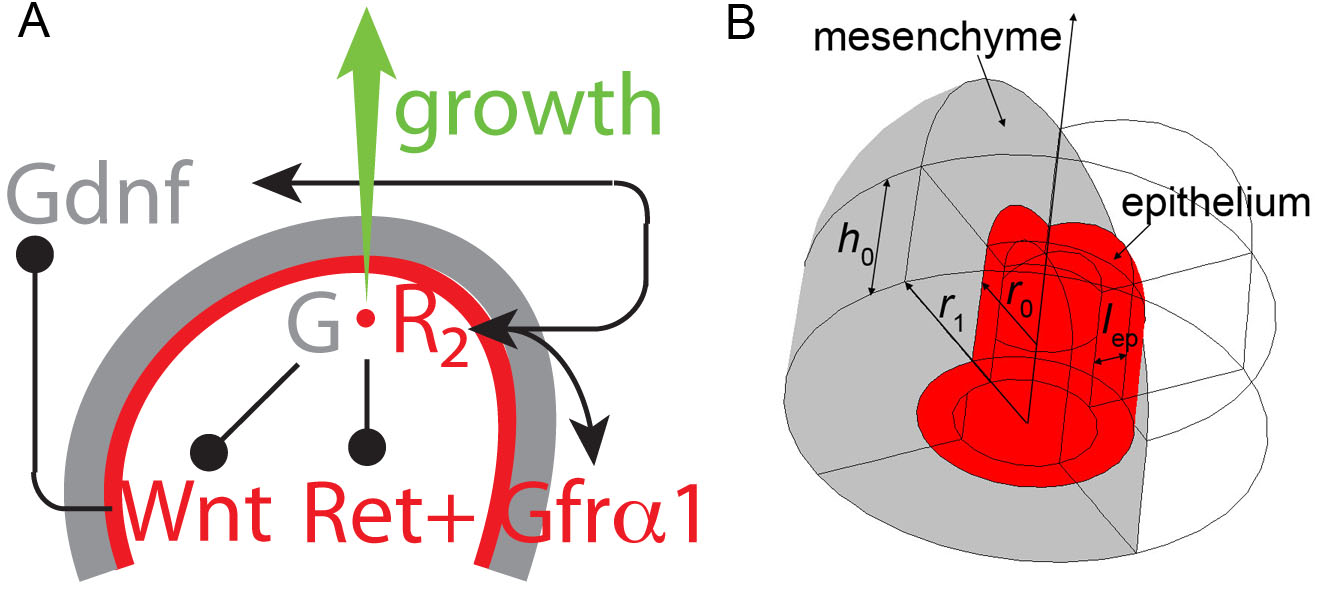}
\caption{\label{fig1} \textbf{Kidney Branching Morphogenesis} \textbf{A:}  The core signalling network that regulates branching in the ureteric bud.  The dimer GDNF ($G$)  binds GFR$\alpha$1 and RET receptor ($R$) to form the  GDNF-RET complex ($GR_2$), which in the model is approximated as a quasi-steady state $GR_2 \sim R^2 G$ . The GDNF-RET complex upregulates expression of the \textit{Ret} receptor and of \textit{Wnt11} ($W$). WNT11 in turn upregulates expression of \textit{Gdnf}. Bud outgrowth is induced by the GDNF-RET signalling. Proteins expressed in the epithelium have their symbol depicted in red, while those expressed in the mesenchyme have their symbols depicted in grey. Arrow-headed lines indicate protein binding interactions, while point-headed lines indicate regulatory interactions.  \textbf{B:} The 3D computational domain comprising epithelium (red) and mesenchyme (grey). The 3D domain is a composite of a cylinder of height $h_0$, inner radius $r_0$, outer radius $r_1$, and epithelial thickness $l_{ep}$.}
\end{center}
\end{figure}

Based on available data we develop and analyse a parsimonous 3-component model for branch point selection during kidney development. We focus on GDNF/RET signalling, as the key driver of branching morphogenesis, as well as on the GDNF/WNT positive feedback loop. The effect of BMP/GREMLIN1 signalling and  of FGF/SPROUTY1 signalling will be discussed, but these components will not be explicitly included in the model. Accordingly the model is restricted to three proteins, GDNF, its receptors (RET and GFR$\alpha$), and WNT11 (Figure \ref{fig1}A). GDNF and WNT11 (which we denote by $G$ and $W$) are secreted proteins. The receptors RET and GFR$\alpha$ (which we denote by $R$)  are membrane bound proteins and therefore diffuse on the surface of a cell with a diffusion coefficient much lower than that of secreted ligands: $\overline{D}_\mathrm{R}<<\overline{D}_\mathrm{W}, \overline{D}_\mathrm{G}$. Earlier studies have successfully described the \emph{in vivo} distribution of morphogens with continuous reaction-diffusion equations on a domain with a length scale as small as 10 cells \cite{Yu2009, Kicheva2007, Ries2009}, and we therefore expect that the ligands in our model are also adequately described by continuous reaction-diffusion equations considering that the number of epithelial cells in any direction is larger than 10. The receptors are more of a concern as these are restricted to the cell surface and their diffusion is thus limited by the cell boundaries in the tissue. We have previously noted that during the receptor half life $t_\mathrm{1/2}=\ln(2)/\delta_R=350~s$ receptors can diffuse over distances of less than the diameter of one epithelial cell  $l=(2D_\mathrm{R}t_{1/2})^{1/2}=3.3~ \mu m$. Moreover, we previously showed numerically that qualitatively similar patterns are observed on a cellurarized lung bud domain \cite{Menshykau:2012kg}. In light of these previous results we also use continuous reaction-diffusion equations for the receptors. 

\textit{Gdnf} is expressed in the metanephric mesenchyme and signals to its receptor pair RET and co-receptor GFR$\alpha$1 (Figure \ref{fig1}A).  The receptor \emph{Ret} and co-receptor \emph{Gfr}$\alpha 1$ of GDNF are both expressed in the epithelium \cite{Pepicelli:1997jz}. The exact stoichiometry and kinetics of the process are still unknown but the GDNF dimer likely first binds to monomeric or dimeric GFR$\alpha$1, and the GDNF-GFR$\alpha$1 complex then interacts with two RET receptors. Some of the RET receptors may be weakly associated with GFR$\alpha$1 before GDNF binding \cite{Eketjall:1999ir}. For simplicity in the first place we only consider one receptor $R$ that represents the rate limiting species, and assume that each GDNF dimer is bound by two such receptor molecules. We note that the model would still work if the number of rate limiting receptor binding events was greater than two or if RET and GFR$\alpha$1 were considered  individually as discussed below. The unbound receptor is lost by degradation at rate $d_\mathrm{R} [\mathrm{R}]$ and by binding to the ligand GDNF at rate $2 d_\mathrm{C} \mathrm{[R]^2} [\mathrm{G}] $. Receptors are constitutively expressed in the epithelium at rate $p_\mathrm{R}$ and their expression is upregulated in the presence of GDNF \cite{Pepicelli:1997jz, Costantini:2010p43730} as described by the term $ p_\mathrm{C} [\mathrm{R}_2\mathrm{G}] $; we note that instead of the linear term we could also use a Hill function with a sufficiently large Hill constant. For simplicity we apply the quasi-steady state approximation to the concentration of the receptor-ligand complex, i.e. $[\mathrm{R}_2\mathrm{G}] = K_\mathrm{C} [\mathrm{R}]^2[\mathrm{G}] $ where $K_C$ is the dissociation constant. We previously showed this approximation to be valid in our lung model for a broad range of physiological parameters  \cite{Menshykau:2012kg}. We then have for the receptor dynamics:

\begin{eqnarray} \label{eq:kidneyR}
\dot{\![\mathrm{R}]} &=& \underbrace{\overline{D}_\mathrm{R} \overline{\Delta} [\mathrm{R}]}_{\text{\ diffusion}} +\underbrace{ p_\mathrm{R}}_{\text{\ production}} \underbrace{-d_{\mathrm{R}}[\mathrm{R}]}_{\text{\ degradation}} \underbrace{+(K_\mathrm{C} p_\mathrm{C} -  2 K_\mathrm{C} d_\mathrm{C}) [\mathrm{R}]^2[\mathrm{G}] }_{\text{\ complex formation $\&$ upregulation}}
\end{eqnarray}

\textit{Gdnf} is expressed in the mesenchyme at rate $\rho_\mathrm{G0}$. Its expression is further enhanced by WNT11 signalling \cite{Majumdar:2003tp, Affolter:2009p25219}. We describe this regulatory interaction by a Hill-type function, $\frac{x^\mathrm{m}}{1+x^\mathrm{m}}$, where $m$ accounts for possible cooperative effects. We use $m=2$ throughout, and we thus have $p_\mathrm{G} \frac{[W]^2}{[W]^2+K_\mathrm{W}}$. GDNF binds to its epithelial receptor at rate $d_\mathrm{C}  [\mathrm{R}]^2[\mathrm{G}] $ and is degraded at rate $-d_{\mathrm{G}}[\mathrm{G}]$. These interactions result in the following mathematical formulation of the GDNF dynamics:

\begin{eqnarray} \label{eq:kidneyG}
\dot{\![\mathrm{G}]} &=& \underbrace{\overline{D}_\mathrm{G} \overline{\Delta} [\mathrm{G}]}_{\text{\ diffusion}} +\underbrace{ p_\mathrm{G0}+p_\mathrm{G} \frac{[W]^2}{[W]^2+K_\mathrm{W}}}_{\text{\ production}} \underbrace{-d_{\mathrm{G}}[\mathrm{G}]}_{\text{\ degradation}}  \underbrace{-K_\mathrm{C}d_\mathrm{C}  [\mathrm{R}]^2[\mathrm{G}] }_{\text{\ complex formation}}
\end{eqnarray}

\emph{Wnt} is expressed  constitutively in the epithelium at rate $p_{W0}$ and its expression is upregulated in response to GDNF-receptor complex formation and signalling, $p_\mathrm{W} \frac{K_\mathrm{C}[\mathrm{R}]^2[\mathrm{G}]}{K_\mathrm{C}[\mathrm{R}]^2[\mathrm{G}]+K_\mathrm{R2G}}$ \cite{Majumdar:2003tp, Affolter:2009p25219}. WNT is lost by degradation at rate $d_\mathrm{W} \mathrm{[W}]$:
\begin{eqnarray} \label{eq:kidneyW}
\dot{\![\mathrm{W}]} &=& \underbrace{\overline{D}_\mathrm{W} \overline{\Delta} [\mathrm{W}]}_{\text{\ diffusion}} + \underbrace{p_\mathrm{W0}+p_\mathrm{W} \frac{K_\mathrm{C}[\mathrm{R}]^2[\mathrm{G}]}{K_\mathrm{C}[\mathrm{R}]^2[\mathrm{G}]+K_\mathrm{R2G}}}_{\text{\ production}} \underbrace{-d_{\mathrm{W}}[\mathrm{W}]}_{\text{\ degradation}} 
\end{eqnarray}

In a separate model we also included the co-receptor GFR$\alpha 1$ explicitly in the model. Its dynamics can be described by 
\begin{equation} \label{eq:kidneyGFR}
\dot{\![\mathrm{GFR}]} = \underbrace{\overline{D}_\mathrm{GFR} \overline{\Delta} [\mathrm{GFR}]}_{\text{\ diffusion}} +\underbrace{ p_\mathrm{GFR}}_{\text{\ production}} \underbrace{-d_{\mathrm{GFR}}[\mathrm{GFR}]}_{\text{\ degradation}} \underbrace{+K_C (p_\mathrm{cGFR} -  d_c) [\mathrm{RET}][\mathrm{GFR}][\mathrm{G}] }_{\text{\ complex formation}} 
\end{equation}
and in \ref{eq:kidneyR}-\ref{eq:kidneyW} we then need to write $[\mathrm{RET}][\mathrm{GFR}][\mathrm{G}]$ instead of  $[\mathrm{RET}]^2[\mathrm{G}]$ for the concentration of the ligand-receptor complex. The model was non-dimensionalized as described in the Methods part to reduce the total number of parameters. The parameters and their values are summarized in Table \ref{tbl:paramD} and a linear stability analysis confirms that these parameter values result in a Turing pattern. To observe the emergence of patterns, simulations were started with no species present, i.e. all concentrations were set to zero at time $t = 0$. 

The idealized computational domain is in the shape of an outgrowing ureteric bud as shown in Figure \ref{fig1}B and comprises two tissue layers, epithelium and mesenchyme. Based on published images we determine the following geometric parameters at the time of branching: internal radius $\overline{r}_0=$50 $\mu m$, epithelial thickness $\overline{l}_{ep} = 15$ $\mu m$, radius of the internal cavity $\overline{r}_0-\overline{l}_\mathrm{ep}$=35 $\mu m$, and radius of the metanephric mesenchyme $\overline{r}_1$=100 $\mu m$. The total domain length $h_0+r_1$ varies and can reach a length of up to 200 $\mu m$   \cite{Muller:1997wl, Lu:2009p43731, Watanabe:2004kr, Chi:2009p43739}. Figures \ref{fig:kidneyDistr}-\ref{fig:mutants} were generated on a static domain with $h_0=0.6$, $r_1 =1$. Figure \ref{fig:UB} was simulated on a growing domain.The implementation of models on a growing domain is discussed in the Methods section.

There is no experimental indication that the mesenchyme and/or epithelium are surrounded by any insulating layer. We therefore assume that the secreted proteins are free to leave the secreting tissue by diffusion. To adequately represent this we embed the ureteric bud in a larger domain into which the ligands can diffuse. The size of this surrounding domain was chosen such that there are no boundary effects; details of the implementation are discussed in the Methods section. Because of the lower computational cost, most of the results presented in this manuscript are calculated with no-flux boundary conditions at the boundary of the domain representing the ureteric bud. However, all  key results were also checked and  reproduced with the ureteric bud embedded into a virtually infinitely large domain, into which the diffusible ligands can diffuse.

Finally, since the length and time scales of the process have been established we can convert the dimensionless diffusion and degradation parameters back to their dimensional counterparts and compare these to experimental values. These converted dimensional parameters are summarized in Table \ref{tbl:paramVal} and all lie well within the experimentally established physiological range.  The protein concentrations have not been established.

\subsection{The GDNF-Receptor Interaction Gives Rise to Modes of Branching Observed During Kidney Morphogenesis}

\begin{figure}[h]
\begin{center}
\includegraphics[width=1\textwidth]{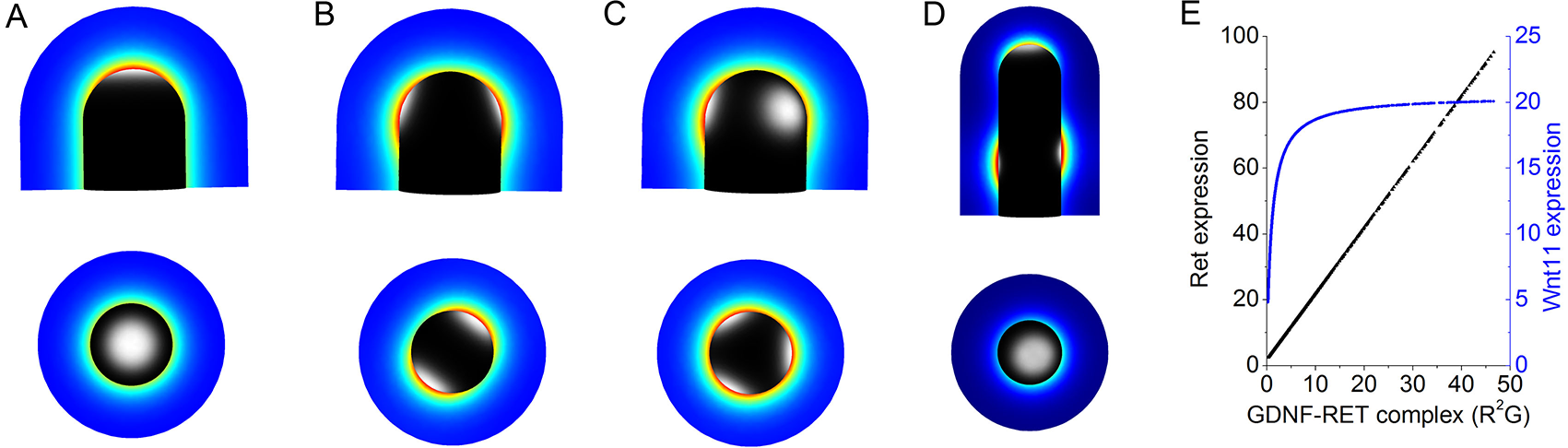}
\caption{\label{fig:kidneyDistr} \textbf{Branching and Gene Expression Patterns}  \textbf{A-D:} The network interactions in Figure \ref{fig1}A result in GDNF-RET receptor signalling patterns ($R^2G$) that correspond to (A) the elongation mode, (B) bifurcations, (C) trifurcations, and (D) lateral branching events. The top panel shows a lateral cut through the 3D bud; the bottom panel shows the top view. The grey scale depicts the concentration of the $R^2G$ signalling complex (white highest, black lowest concentration) at the border of epithelium and mesenchyme. The GDNF-RET receptor signalling complex lies adjacent to Gdnf expressing zones in the mesenchyme (indicated in rainbow color code with red indicating  highest, blue indicating lowest expression levels). The simulations were carried out with  the parameter values in Table \ref{tbl:paramD} except for  (A) where $v=1.5$, and  (B) where $v=2$. Panel D was calculated with a model which explicitly considers receptor and co-receptor. Parameter values were used as in the Table \ref{tbl:paramD}, except for $h_0=2$, $v_1=1.15$ and $\rho_{R1}=1$. Index 1 indicates that parameter values where altered only for the receptor (or the co-receptor) while parameter values for the other where kept as in Table \ref{tbl:paramD}. Initial model has identical parameter values for the receptor and co-receptor, so we cannot say if change parameter for the receptor of for the co-receptor. \textbf{E:} The level of Ret expression (black line) linearly correlates with the activity of the GDNF-RET receptor complex, $R^2G$,  while the expression level of Wnt11 (blue line) non-linearly positively correlates with $R^2G$. The panel reports the measured expression levels of Ret and Wnt11 versus the concentration of the GDNF-RET receptor complex ($R^2G$) at each mesh point.  }
\end{center}
\end{figure}

Epithelial outgrowth is induced by GDNF signalling \cite{Tang:2002it,Tang:1998vf,Sariola:2003jo,Costantini:2010p43730, Majumdar:2003tp, Treanor:1996dq, Pichel:1996en, Pepicelli:1997jz, Sanchez:1996cy}. The only known GDNF receptor which transmits the GDNF signal into the cell is RET, with its co-receptor GFR$\alpha$1 \cite{Costantini:2006p43749, Costantini:2010p43730}. We therefore use the GDNF-RET complex concentration, $R^2G$, as a marker for the points of bud outgrowth. When we solve the model (equations \ref{eq:kidneyD}) for the signalling interactions depicted in Figure \ref{fig1}A on the 3D computational domain depicted in Figure \ref{fig1}B we observe distributions of the GDNF-receptor complex, $R^2G$, that can in principle explain the different branching modes observed in the kidney (Figure  \ref{fig:kidneyDistr}A-D, grey scale with white indicating the highest concentrations). Since quantitative data on absolute protein concentrations and expression levels are not available we have to restrict ourselves to a qualitative discussion of distribution patterns. We therefore deliberately left out scale bars for better readability. Depending on the choice of parameters we observe GDNF-receptor complexes concentrated either at the tip (elongation mode, Figure  \ref{fig:kidneyDistr}A), in two spots at the side (bifurcation mode, Figure  \ref{fig:kidneyDistr}B), in three spots on the side (trifurcation mode, Figure  \ref{fig:kidneyDistr}C), or at the tip and in spots on the side (lateral branching mode, Figure  \ref{fig:kidneyDistr}D). We note that small changes in the parameter values were sufficient to switch between the different branching modes, i.e. $v=2.5$ for the trifurcation mode was changed to $v=1.5$ to obtain the elongation mode in panel A, and to $v=2$ to obtain the bifurcating pattern in panel B. The lateral branching pattern was obtained with a model that considered both the receptor RET and its co-receptor GFR$\alpha$1 explicitly, and required as parameter values $h_0 = 2$, $v_1 = 1.15$, and $\rho_{R1} = 1$. The implications of such sensitivity on both, pattern robustness and regulatory potential of other factors, is discussed below.

As part of the regulatory mechanism shown in Figure \ref{fig1}A signalling by the GDNF-RET receptor complex induces expression of \textit{Ret} and \textit{Wnt11}. As expected from the model formulation, the expression of these two genes in the bud domain is indeed positively correlated with the level of GDNF-RET signalling. Thus, when we measured the expression levels of \textit{Ret} and \textit{Wnt11} expression as well as the concentration of the GDNF-RET receptor complex ($R^2G$) at each mesh point we found that these are positively correlated (Figure  \ref{fig:kidneyDistr}E). The co-localisation of the strongest levels of receptor and \textit{Wnt11} expression with the GDNF-receptor complex agrees well with the experimentally observed expression patterns \cite{Majumdar:2003tp}. WNT11 signalling in turn upregulates the expression of \textit{Gdnf} in the mesenchyme. The \textit{Gdnf} expressing zones are indeed adjacent to the epithelial GDNF-RET receptor signalling patches (\textit{Gdnf} expression levels are shown in rainbow colour code in Figure  \ref{fig:kidneyDistr}A-D).

\subsection{Branch Mode Preferences in Lung and Kidney}

\begin{figure}[h]
\begin{center}
\includegraphics[width=0.12\textwidth]{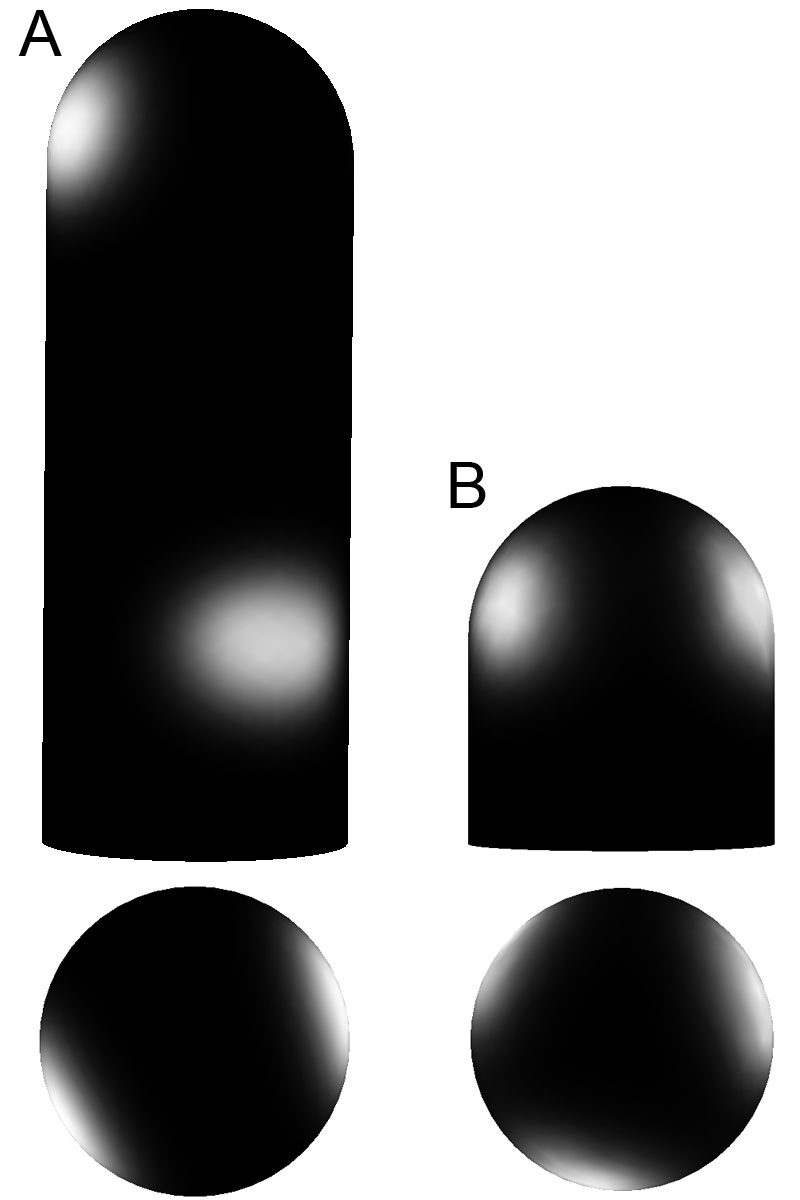}
\caption{\label{fig:lateral} \textbf{The  Lateral Branching patterns are highly sensitive to parameter variations.} \textbf{A:} The GDNF-RET receptor  ($R^2G$) pattern that correspond to the lateral branching pattern is not robust to the addition of 5\% Gaussian noise (i.e. relative standard deviation of 0.05) to the parameter values. Calculations were carried out with the model from Figure \ref{fig:kidneyDistr}D. Note that the spot at the tip vanishes upon addition of 5\% Gaussian noise. \textbf{B:} The trifurcation branching pattern is robust even to the addition of 50\% Gaussian noise to the parameter values. The grey scale depicts the concentration of the $R^2G$ signalling complex in the epithelium (white highest, black lowest concentration); the mesenchyme is not shown, but was included in the simulations.} 
\end{center}
\end{figure}

While the same types of branching events have been observed in lungs and kidneys their frequency of use differs greatly. Thus, lateral branching has been found mainly in the lung, but rarely in the kidney, while trifurcations are rare in the lung, but  more prevalent in the kidney \cite{Metzger:2008ky, Blanc:2012ea,Watanabe:2004kr}. The dominant branching mode in the ureteric bud is terminal bifurcations \cite{Watanabe:2004kr}. Similarly, in our models we observe patterns that correspond to all observed modes of branching (Figure  \ref{fig:kidneyDistr}A-D), but their frequency differs. Thus, in our previous model for the lung we observed lateral branching events for a large part of the parameter space \cite{Menshykau:2012kg}, while in the model for the ureteric bud lateral branches are observed only rarely, and if observed, they are very sensitive to changes in the parameter values (Figure \ref{fig:lateral}). Thus, adding as little as 5$\%$ Gaussian noise to the parameter values used for Figure 2D caused the pattern to change away from the lateral branching mode as shown for one example in Figure \ref{fig:lateral}A, where the spot at the tip of the bud (that would support further bud elongation during lateral domain branching) vanished. On the contrary we could add 50$\%$ noise to the parameter values without affecting the trifurcation pattern (Figure \ref{fig:lateral}B). The kidney model would thus predict lateral branching events to be rare, while bifurcations and trifurcations are observed for a wide range of parameter values (Figure \ref{fig:mutants}A). While we cannot exclude that these pattern preferences are the result of our particular parameter choices in the two models for the lung and kidney, we note that the particular choice of parameter values is heavily constrained by the mutant phenotypes. The models for the lung and the kidney may appear mathematically similar at first sight but there are important differences. For one the FGF10/SHH interaction results in a negative feedback while the GDNF/WNT11 interaction results in a positive feedback. Furthermore, in case of the ureteric bud the outgrowth inducing ligand  (GDNF) is produced in the mesenchyme and its receptor (RET/GFR$\alpha$) in the epithelium, while in our lung model the outgrowth-inhibiting factor (SHH) is produced in the epithelium and its receptor (PTCH1) in the mesenchyme. However, the latter difference may be less significant as a similar Schnakenberg-type Turing model for the lung could also be constructed based on the FGF10-receptor interaction. In that case, much as in the ureteric bud, the outgrowth-inducing factor, FGF10, would be produced in the mesenchyme. Such alternative model has not yet been analysed by us in detail and we can therefore not comment on the branch type preference. 

To further establish the size of the parameter space for which we obtain bifurcations, trifurcations, and elongation patterns we varied each parameter one by one until we observed a mode change. As can be seen from Figure \ref{fig:mutants}A trifurcation patterns are the most robust patterns. At the same time almost each parameter can be employed to switch the pattern between bifurcation and trifurcation modes of branching. Trifurcations and bifurcations thus appears to be robust to small variations in parameter values, yet sensitive to regulation. The many additional interactions that have been described, but that were not included in this parsimonious model thus can exhibit their documented impact on branching morphogenesis by altering one or several of these parameters. Thus, FGF signalling appears to control similar downstream targets as RET signalling \cite{Michos:2010p43732} and \textit{Fgfr2}-mutant kidneys exhibit reduced branching as observed in simulations with less \textit{Ret} expression (Figure \ref{fig:mutants}B).  BMP signalling appears to antagonize WNT, and addition of high concentrations of BMPs inhibits branching and removal of the BMP antagonist \textit{Grem1} inhibits ureteric bud outgrowth from the Wolffian duct  \cite{Bush:2004uc, Clark:2001gp, Piscione:1997uba, Liu:2006p5513, Xu:2011ka, Michos:2007p42720}. Likewise in the simulations lower levels of $W$ production results in less branching (Figure \ref{fig:mutants}B).

Interestingly, a model where RET and GFR$\alpha$ are considered individually is more robust to parameter variations than the simplified model with only one receptor, $R$ (Figure \ref{fig:mutants}A). Thus, assembly of a complex from individually regulated subunits increases the robustness to variations in the concentrations of the subunits.

\subsection{Mutants}
An important test for the suitability of a mathematical model is its consistency with a wide range of independent experimental observations. Mutants represent a perturbation of the parameter set of the original system behavior and if a wide range of such perturbations can be reproduced correctly by the model then this provides strong support in favor of a model. Below we discuss the relevant mutants reported in the literature. To compare \emph{in silico} mutants with those generated \emph{in vivo} we used the following read-outs: the \textit{Gdnf} expression level (Figure \ref{fig:mutants}B), the GDNF-receptor signalling activity (Figure \ref{fig:mutants}C), and the extent of branching (Figure \ref{fig:mutants}D). GDNF is the main signalling protein controlling ureteric mesenchyme outgrowth and GDNF signals through its receptor RET and co-receptor GFR$\alpha$. Accordingly, the GDNF-receptor activity was computed as the total concentration of $R^2G$ in the epithelium. GDNF signalling induces \textit{Wnt} and \textit{Ret} expression, and GDNF signalling levels therefore also serve as a proxy for the \textit{Wnt} and \textit{Ret} expression levels. The extent of branching was measured via the frequency of the Turing pattern which is inversely proportional to the distance between two spots. Smaller values of the frequency therefore correspond to kidneys with reduced branching and longer branches. Such correlation between the extent of branching and branch length is indeed observed in embryos. Thus, the \textit{Tgf}$\beta2^{+/-}$ mutant exhibits reduced branching and branches are of increased length \cite{Short:2010p49323}. 

Homozygous mutants were implemented by setting the protein production rates to zero (e.g. $\rho_{W0}$ and $\rho_{W}$ in case of \emph{Wnt11}). Heterozygous mutants retain one active copy of the gene and accordingly the production rates were halved, even though we acknowledge that net changes may be different in mutants because of further feedbacks that were not considered in the model. 

\subsubsection{\textit{Gdnf} Mutants} GDNF is an important ligand controlling kidney branching morphogenesis, and in the \emph{Gdnf}$^{-/-}$ mutant the ureteric epithelium fails to invade the mesenchyme and no kidney forms \cite{Moore:1996io, Costantini:2006p43749, Pichel:1996en}. In our computational model GDNF is part of the core patterning mechanism and in its absence no pattern or branching is observed. \emph{Gdnf}$^{+/-}$ mutants have reduced branching and reduced expression levels of \emph{Gdnf}, \emph{Ret} and \emph{Wnt11} \cite{Pichel:1996en, Pepicelli:1997jz}. The computational model reproduces these mutants and exhibits reduced \textit{Gdnf} expression levels (Figure \ref{fig:mutants}B), reduced GDNF signalling (Figure \ref{fig:mutants}C) (which correlates with the \textit{Ret} and \textit{Wnt} expression levels (Figure \ref{fig:kidneyDistr}), and an increased wavelength of the Turing pattern (Figure \ref{fig:mutants}D), as characteristic for reduced branching. The \textit{Gdnf} expression levels are below 50\% in the simulations because of the lower levels of positive WNT feedback in the heterozygous mutant.

\subsubsection{\textit{Ret} and \emph{Gfr}$\alpha$ Mutants}
\emph{Ret}$^{-/-}$ and \emph{Gfr}$\alpha^{-/-}$ mutant mice fail to develop kidneys \cite{Costantini:2006p43749, Schuchardt:1994hg, Enomoto1998} while \emph{Ret}$^{+/-}$ mutant mice develop kidneys of normal size. In our computational model the receptors RET and GFR$\alpha$1 are part of the core patterning mechanism and in their absence no pattern or branching is observed. The model predicts for $Ret^{+/-}$ mutants a small decrease in the expression and signalling levels (Figure \ref{fig:mutants}B,C) as well as slightly reduced branching (Figure \ref{fig:mutants}D). Experiments show that heterozygous \emph{Ret} mutants demonstrate normal \cite{Majumdar:2003tp} or slightly reduced branching \cite{Clevers2006} (Figure \ref{fig:mutants} E). Note that \textit{Ret} and \emph{Gfr}$\alpha$ mutants are computed according to a model, which considers explicitly both RET and GFR$\alpha 1$ receptors (Equation \ref{eq:kidneyDSUP}). 

\subsubsection{\textit{Wnt11} mutants}
Unlike \emph{Gdnf}$^{-/-}$, \emph{Ret}$^{-/-}$ and \emph{Gfr}$\alpha^{-/-}$ mutants, \emph{Wnt11}$^{-/-}$ mice develop kidneys. The number of branches and the organ size are, however, reduced \cite{Majumdar:2003tp}. WNT signalling is transmitted via the receptor Frizzled (FZ); multiple receptors of the family are known to transduce WNT signals  \cite{Bhanot1996, Logan2004, Clevers2006}. In cell culture, FZ4 and FZ8 can mediate noncanonical signalling stimulated by WNT11, but only FZ4 mediates WNT11-stimulated canonical signalling \cite{Ye:2011fz}. The   \emph{Wnt11}$^{-/-}$ and  \emph{Fz4}$^{-/-}$/ \emph{Fz8}$^{-/-}$ mutants have similar kidney phenotypes  \cite{Ye:2011fz}.

\begin{figure}[h]
\begin{center}
\vspace{-2cm}
\includegraphics[width=1\textwidth]{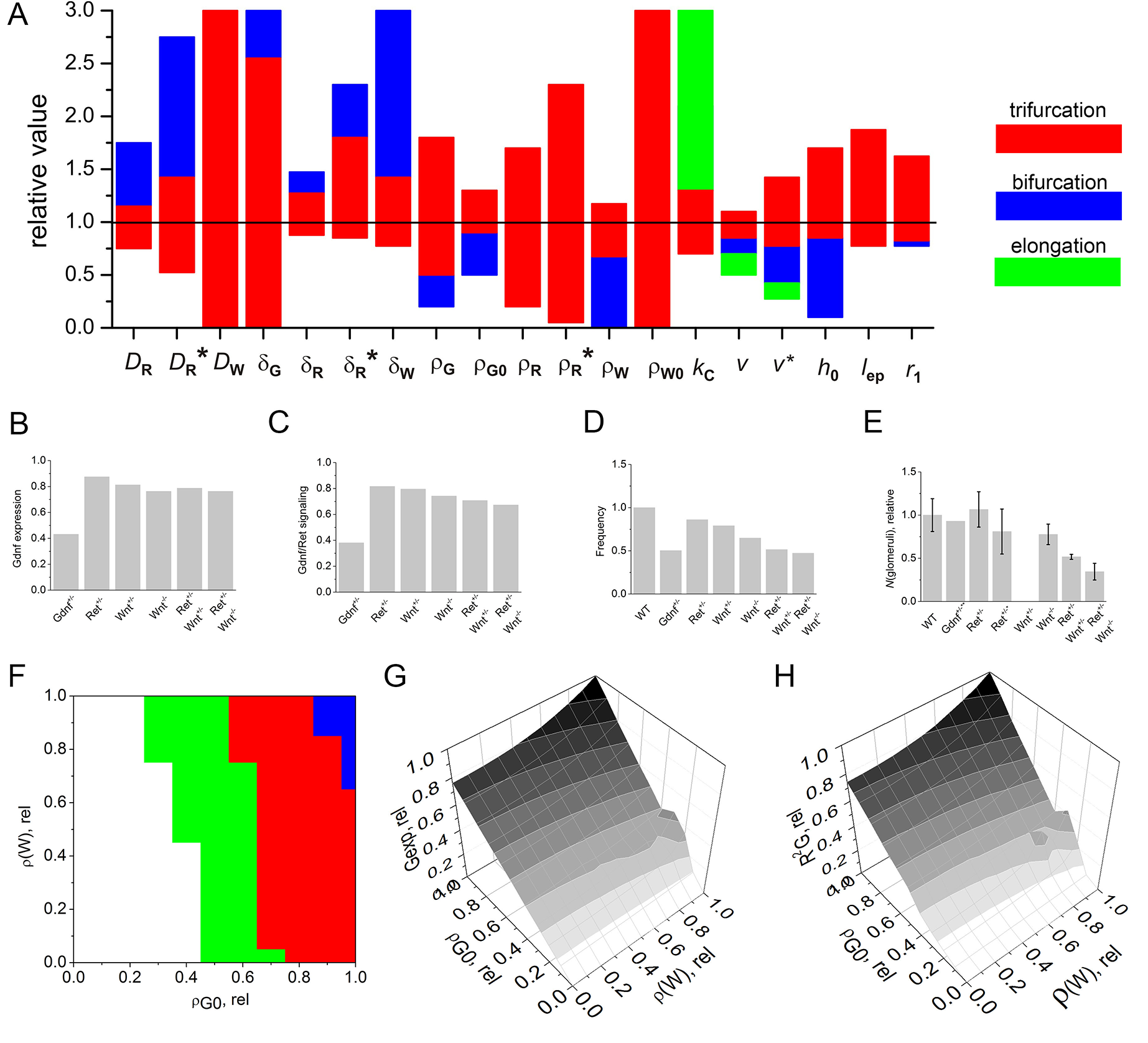}
\caption{\label{fig:mutants} \textbf{Sensitivity to Parameter Changes and Predictions of Mutant Phenotypes. } 
 \textbf{A:} Sensitivity to perturbations in single parameter values. Parameter values from Table \ref{tbl:paramD}  were varied one by one and the pattern of the GDNF-RET receptor complex, $R^2G$, was evaluated with respect to the branching modes (trifurcation -red, bifurcation-blue, elongation - green) as defined in Figure \ref{fig:kidneyDistr}A-D. The lateral branching mode was not observed for this parameter set. Parameters marked with a star (*) were evaluated in a model which explicitly considered both receptor RET and co-receptor GFR$\alpha$1. The accuracy of the borders of the parameter ranges was 5$\%$ or higher. The interpretation of parameters are presented in Table \ref{tbl:paramD}.  \textbf{B-D:} Predicted mutant phenotypes.  \textbf{B:} The level of Gdnf expression, \textbf{C:} the level of GDNF signalling (i.e. the GDNF-RET receptor concentration $R^2G$), and \textbf{D:} and the extent of branching (as evaluated based on the pattern wavelength)  in the indicated mutants relative to WT expression levels. The parameter values in Table 1 define the WT condition. \textbf{E:} Relative number of counted glomeruli as reproduced from \cite{Majumdar:2003tp}, except for the $\textit{Ret}^{+/-*}$ mutant, which is reproduced from \cite{Clarke2006}, the $\textit{Gdnf}^{+/-**}$ mutant, which is reproduced from \cite{Michos:2010p43732} and the $Gdnf^{+/-***}$ mutant, which is reproduced from \cite{Pichel:1996en}. \textbf{F-H:} The predicted impact of combined relative changes in the WNT-independent \textit{Gdnf} expression rate $\rho_{G0}$ and in the \textit{Wnt} expression rate $\rho_{W}$ on (F) the branching mode, (G) \textit{Gdnf} expression, and (H) the concentration of the GDNF-RET complex ($R^2G$) integrated over the entire epithelium. Wildtype rates correspond to 1. }
\end{center}
\end{figure}

To simulate the \emph{Wnt11}$^{-/-}$ mutants we set $\rho_\mathrm{W}$ and $\rho_\mathrm{W0}$ to zero. Much as in the embryo, the extent of branching is reduced in \textit{Wnt11} mutants (Figure \ref{fig:mutants}D). Unlike in the lung where the distance between buds increases abruptly as the \textit{Fgf10} expression levels fall below a threshold in the \textit{Fgf10} allelic sequence \cite{Ramasamy:2007hu}, there is a gradual decrease of branching in the \textit{Ret/Wnt11} allelic sequence (\ref{fig:mutants}E) \cite{Majumdar:2003tp}. Both the lung \cite{Menshykau:2012kg} and the kidney phenotype (Figure \ref{fig:mutants}D) are reproduced by the respective models. 

The model considers only one WNT ligand, WNT11, which enhances \textit{Gdnf} expression. The other WNTs are likely to also enhance \textit{Gdnf} expression, an effect that is incorporated implicitly in the WNT11-independent \textit{Gdnf} expression rate. We find that reducing either only the \textit{Wnt11} expression rate $\rho_{W}$, only the WNT11-independent \textit{Gdnf} expression rate $\rho_{G0}$, or both can all alter the branching mode in a synergistic manner (Figure \ref{fig:mutants}F). We further find that setting  $\rho_{W}$ to zero reduces the \textit{Gdnf} expression (Figure \ref{fig:mutants}G) and the GDNF-receptor complex concentration by about 20\% (Figure \ref{fig:mutants}H). A parallel reduction in $\rho_{W}$ and $\rho_{G0}$ further reduces \textit{Gdnf} expression, such that for a 20\% reduction in $\rho_{G0}$ and $\rho_{W}$ \textit{Gdnf} expression and GDNF-receptor complex concentration are reduced to about 50\% (Figure \ref{fig:mutants}G,H).

\subsection{Ureteric Bud Outgrowth}
\begin{figure}[h]
\begin{center}
\includegraphics[width=1\textwidth]{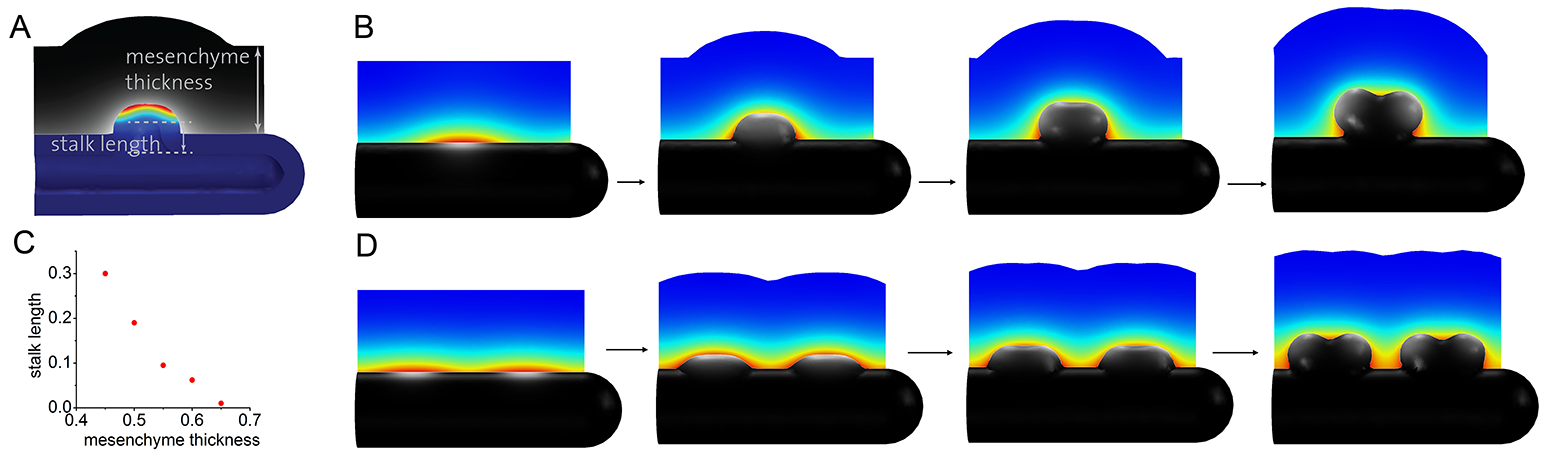}
\caption{\label{fig:UB} \textbf{Branching of the Ureteric Bud into the Metanephric Mesenchyme. }   \textbf{A:} The computational domain for the simulation of an ureteric bud branching from the Wolffian duct into the metanephric mesenchyme. \textbf{B:} The branching of the ureteric bud into the metanephric mesenchyme. The grey scale depicts the concentration of the $R^2G$ signalling complex (white highest, black lowest concentration) at the border of epithelium and mesenchyme. The GDNF-RET receptor signalling complex lies adjacent to \textit{Gdnf} expressing zones in the mesenchyme (indicated in rainbow color code with red indicating  highest, blue indicating lowest expression levels).  \textbf{C:} The dependency of the non-dimensional stalk length as defined in panel A on the thickness of the mesenchyme. \textbf{D:} The branching of the ureteric bud into the metanephric mesenchyme in a \textit{Sprouty}$^{-/-}$ mutant.  The parameter values are as in Table 1, except $\rho_\mathrm{G0}$=0.38. Colour code as in panel B.}
\end{center}
\end{figure}

Genetic analysis shows that the network that controls mesenchyme invasion by  the ureteric bud is similar to the one controlling branching \cite{Costantini:2010p43730}. The ureteric buds of homozygous \textit{Gdnf}, \textit{Ret} or \textit{Gfr$\alpha$1} knockouts fail to invade the metanephric mesenchyme \cite{Costantini:2010p43730, Majumdar:2003tp, Treanor:1996dq, Pichel:1996en, Pepicelli:1997jz, Sanchez:1996cy}. WNT11 seems to be less important because an invasion and a first round of branching are still observed in homozygous \textit{Wnt11} knockouts \cite{Majumdar:2003tp}. Figure \ref{fig:UB}A shows the computational domain that was used to represent the Wolffian duct, from which the ureteric bud branches out. The core regulatory network depicted in Figure \ref{fig1}A is sufficient to control ureteric bud outgrowth and a first round of branching  \emph{in silico} (Figure \ref{fig:UB}B). Recently it was reported that the thickness of the mesenchyme of the embryonic lung plays an important role during branching morphogenesis, in that it has an impact on the outgrowth of the epithelium \cite{Blanc:2012ea}. The thickness of the mesenchyme is indeed also important in our simulations of kidney branching morphogenesis: the thicker the mesenchyme, the shorter the stalk, and the earlier the first round of branching is observed (Figure \ref{fig:UB}C). If the mesenchyme thickness exceeds some threshold value then the mesenchyme will be invaded in more than one place, similar to the case shown in Figure \ref{fig:UB}D. The impact of the mesenchyme thickness can be rationalized in the following way: the dimensionless diffusional length scale of GDNF in our model is $l\sim(2D_\mathrm{G}\ln(2)/\delta_{G})^{1/2}=3.5$. The thickness of the mesenchyme is around 0.5. Therefore, the thicker the mesenchyme the more GDNF can be supplied to the epithelium-mesenchyme border where GDNF binds to its receptors. An increased mesenchyme thickness has therefore an effect similar to an enhanced GDNF production rate. 

Interestingly, mutants with reduced expression levels of the  inhibitor of GDNF signalling, \textit{Sprouty}, develop multiple kidneys \cite{Basson2005, Michos:2010p43732}. Removal of an inhibitor of GDNF signalling corresponds to a reduction of the signalling threshold $K_{R2G}$ in our model. In the non-dimensional model a reduction in $K_{R2G}$ corresponds to an increase in the \textit{Ret} and \textit{Gdnf} production rates, $\rho_R$, $\rho_{G0}$, and $\rho_G$ respectively (Equation \ref{eq:kidneyD}).To address the \textit{Sprouty} mutants in our simulation we increased the production rate of \textit{Gdnf}, $\rho_\mathrm{G0}$. Much as in the experiments, the simulations with an increased production rate of \textit{Gdnf} indeed resulted in the invasion of the mesenchyme by the epithelium in several places (Figure \ref{fig:UB}D). Both the thicker mesenchyme and the removal of \textit{Sprouty} thus correspond to increased \textit{Gdnf} production in our non-dimensional model. Sprouty has not been included into the model explicitly because it is an intracellular protein, while the model is focused on spatio-temporal patterning and therefore includes only secreted ligands and their receptors.

\section{Discussion}

While many regulatory components and local interactions have been defined an integrated understanding of the regulatory networks that control the different branching processes is lacking. We have shown here that the biochemical interactions between the core regulatory factor GDNF and its receptor result in a Schnakenberg-type Turing mechanism. The model is similar to the one that we previously obtained for the interaction of SHH and PTCH1 in the developing lung  \cite{Menshykau:2012kg}. An important difference lies in the feedback architecture in that GDNF and WNT11 engage in a positive feedback while SHH forms a negative feedback with FGF10 in the lung.  Moreover, while the diffusible ligands and receptors are expressed in separate tissue layers in both developing organs, \textit{Shh} is expressed in the epithelium while \textit{Gdnf} is expressed in the mesenchyme. Interestingly, while both mechanisms give rise to the full range of patterns that would correspond to the observed branching types (lateral branching, bifurcations, trifurcations) the GDNF/WNT11 based mechanism favours bifurcations and trifurcations while the  SHH/FGF10 mechanism favours lateral branching events. This is in good agreement with the observed branching patterns in the two organs. 

Unlike our model for lung development we solved the kidney model on growing domains that deformed in 3D  in response to the local GDNF-receptor concentration. As the domain length increased  further, patterns emerge, as is expected for a Turing mechanism. We further noticed that not only the length of the bud, but also the thickness of the mesenchyme affects the distance between two branching events in that the  length of the outgrowing stalk is shorter and branching starts earlier if the mesenchyme is thicker. This is in good agreement with experimental observations in the lung where side-branching was noted to occur when sufficient space becomes available around the circumference of a parent branch  \cite{Blanc:2012ea}.

It is so far unknown how the sequence of branching events is controlled during organogenesis, but given the stereotyped nature of early branching events, stochastic effects are unlikely to play a role and the branching sequence should emerge from the patterning mechanism in a deterministic fashion. We have previously shown in our model of lung branching morphogenesis that the signalling spots that emerge during bud outgrowth do not all have the same intensity \cite{Menshykau:2012kg}. We further showed that the growth speed can affect the branch mode \cite{Menshykau:2012kg}. If the growth speed depended on the strength of the signal then this could, in principle, result in a sequence of different branching events during bud outgrowth. In this manuscript we also show that the thickness of mesenchyme can influence branch mode selection. As kidney branching progresses the amount of mesenchyme surrounding the epithelial tip changes. In principle, this could influence branch mode selection. More work is clearly needed, both experimentally and computationally, to address the mechanism of branch mode selection.

Turing mechanisms have been proposed for many other biological patterning phenomena, and reproduce the size and geometry-dependence of biological patterns of various complexity \cite{Kondo:2010bx,MurrayBook}. However, it has remained difficult to firmly establish their use in biological pattern formation \cite{Hoefer:tr}, and in several cases Turing-type mechanisms have been wrongly assigned to patterning processes such as e.g. the mechanism by which the stripy expression pattern of pair-rule genes emerge during Drosophila development \cite{Akam:1989kk}. These failures reveal the importance of a careful and comprehensive analysis of the underlying molecular interactions before proposing a Turing mechanism. The emerging Turing patterns are highly sensitive to the particular type of biochemical interactions and the parameter values.  This aspect can be used to thoroughly test proposed Turing mechanisms with data from mutants. We have previously shown that the SHH/FGF10-based Turing mechanisms recapitulates even counterintuitive mutant phenotypes such as the abrupt increase in the spacings between buds in the \textit{Fgf10} allelic sequence as the \textit{Fgf10} expression levels fall below a threshold  \cite{Menshykau:2012kg}, and we have shown here that also the GDNF/RET-based Turing mechanism recapitulates all observed mutant phenotypes in the ureteric bud, including the gradual decrease of branching in the \textit{Ret/Wnt11} allelic sequence.

A number of alternative models have been proposed to explain the control of branching in other organs, that focused mainly on geometric rather than signalling effects, i.e. \cite{Nelson:2006gn, Hirashima:2009p43515, Miura:2008p43508, Lubkin:2008iy}. It is likely that both geometric and signalling effects work together to establish the observed patterns, but it is to be expected that signalling presents the dominating regulatory control. 

In summary, Turing mechanisms offer a reliable mechanism for symmetry breaking that allows the outgrowing bud to change from its approximately cylindrical symmetry to the rotational symmetry after undergoing branching. The various branching events, i.e. bifurcations, trifurcation or lateral branching, can all result from the same regulatory interactions, thus permitting the implementation of complex branching pattern with a limited number of proteins. Different feedback architectures can modulate the branching type and sequence.  At the same time different signalling systems can be used to implement such ligand-receptor based Turing mechanisms, such that branching in the lung can be explained with a receptor-ligand based Turing mechanism based on FGF10 and / or SHH, and branching in the ureteric bud can be explained with receptor-ligand based Turing mechanism based on GDNF. Unlike classical activator-inhibitor Turing mechanisms, ligand-receptor-based Turing mechanisms can be implemented by a single ligand, as long as the ligand-receptor interaction is cooperative, ligand-receptor binding results in an increased emergence of receptors on the membrane, and ligand diffuses faster than its receptor. This makes them a versatile mechanism for spontaneous pattern formation during development. 

\section{Methods}
\subsection{Non-Dimensionalization of the Model}
\ref{eq:kidneyR}-\ref{eq:kidneyW} were non-dimensionalized to reduce the total number of parameters:
\begin{eqnarray} \label{eq:kidneyD}
\dot{G} &=&  \Delta  G +  \rho_\mathrm{G0} + \rho_\mathrm{G} \frac{W^2}{W^2 + 1} - \delta_\mathrm{G} G - \delta_\mathrm{C} R^2 G \nonumber\\
\dot{R} &=& D_\mathrm{R} \Delta R +  \rho_\mathrm{R}  + (\nu- 2\delta_\mathrm{C})  R^2 G   - \delta_\mathrm{R} R   \nonumber\\
\dot{W} &=& D_\mathrm{W}  \Delta  W +    \rho_{W0} +\rho_\mathrm{W}   \frac{R^2 G}{R^2 G + 1}- \delta_\mathrm{W}W
\end{eqnarray}
where, $\tau=t \overline{D}_\mathrm{G}/r_1^2$; $D_\mathrm{W}= \overline{D}_W/\overline{D}_\mathrm{G}$, $D_\mathrm{R} = \overline{D}_\mathrm{R}/\overline{D}_\mathrm{G}$; $W = [W] / K_\mathrm{W}$, $R = [R]  K_\mathrm{C}^{1/3}K_\mathrm{R2G}^{-1/3}$; $G = [G]  K_\mathrm{C}^{1/3}K_\mathrm{R2G}^{-1/3}$; $\delta_\mathrm{i}$=$d_\mathrm{i}r_1^2/\overline{D}_\mathrm{G}$ where $i=R,G,W$; $\rho_\mathrm{W0} = p_\mathrm{W0} r_1^2 /({K_\mathrm{W} \overline{D}_\mathrm{G}})$; $\rho_\mathrm{W} = p_\mathrm{W} r_1^2 /({K_\mathrm{W} \overline{D}_\mathrm{G}})$; $\rho_\mathrm{i} = p_\mathrm{i}  {K_\mathrm{C}^{1/3}K_\mathrm{R2G}^{-1/3}  r_1^2 /\overline{D}_\mathrm{G}}$ where $i=R, G, G_0$. \\

\noindent The non-dimensional equation for the co-receptor GFR$\alpha 1$ reads
\begin{eqnarray} \label{eq:kidneyDSUP}
\dot{GFR} &=&  \Delta GFR +  \rho_\mathrm{GFR}   - \delta_\mathrm{GFR} G + (\nu_\mathrm{GFR}- \delta_\mathrm{C})  R\times GFR\times G 
\end{eqnarray}
and in \ref{eq:kidneyD} we then need to write $R\times G\times G$ instead of  $R^2 \times G$ for the concentration of the ligand-receptor complex. The parameters and their values are summarized in Table \ref{tbl:paramD}. In the limit of $W^2>>1$ and $W^2<<1$ the partial differential equations (PDEs) that describe the dynamics of the variables $G$ and $R$ are independent of the variable $W$ and the system of equations \ref{eq:kidneyD} reduces to the Schnakenberg model.  \\

\subsection{Linear Stability Analysis}
To confirm that the observed patterns result from a Turing mechanism we performed a linear stability analysis at the steady state for the parameter values in Table \ref{tbl:paramD}. To that end we determined the eigenvalues of the Jacobian $J$ in the absence and presence of diffusion and show that all real parts of the eigenvalues are negative in the absence of diffusion so that the steady state is stable in the absence of diffusion. In the presence of diffusion we obtain at least on eigenvalue with positive real part such that we have a diffusion-driven instability. The Jacobian for the PDE system given by  \ref{eq:kidneyD} at the steady-state (G=0.434, R=3.13, W=34.4) in the absence of diffusion is given by: 
\begin{eqnarray}
J = \left(
\begin{array}{ccc}
 -R^2 \text{$\delta $c}-\text{$\delta $g} & -2 G R \text{$\delta $c} & -\frac{2 W^3 \text{$\rho $g}}{\left(1+W^2\right)^2}+\frac{2 W \text{$\rho $g}}{1+W^2} \\
 R^2 (v-2 \text{$\delta $c}) & 2 G R (v-2 \text{$\delta $c})-\text{$\delta $r} & 0 \\
 -\frac{G R^4 \text{$\rho $w}}{\left(1+G R^2\right)^2}+\frac{R^2 \text{$\rho $w}}{1+G R^2} & -\frac{2 G^2 R^3 \text{$\rho $w}}{\left(1+G R^2\right)^2}+\frac{2 G R \text{$\rho $w}}{1+G R^2} & -\text{$\delta $w}
\end{array}
\right)
&=&\nonumber \\
\left(
\begin{array}{ccc}
 -5.01 & -1.36 & 9.80\times10^{-5} \\
 9.83 & 0.724 & 0 \\
 7.08 & 1.96 & -0.5
\end{array}
\right)
\end{eqnarray}
All eigenvalues of this Jacobian have a negative real part: $\lambda_{1,2}=-2.15\pm2.27i$, $\lambda_3=-0.500$. In the presence of diffusion the Jacobian
\begin{eqnarray}
J = \left(
\begin{array}{ccc}
 -D_\mathrm{G} k^2-R^2 \text{$\delta $c}-\text{$\delta $g}  & -2 G R \text{$\delta $c} & -\frac{2 W^3 \text{$\rho $g}}{\left(1+W^2\right)^2}+\frac{2 W \text{$\rho $g}}{1+W^2} \\
 R^2 (v-2 \text{$\delta $c}) & -D_\mathrm{R}  k^2+2 G R (v-2 \text{$\delta $c})-\text{$\delta $r}  & 0 \\
 -\frac{G R^4 \text{$\rho $w}}{\left(1+G R^2\right)^2}+\frac{R^2 \text{$\rho $w}}{1+G R^2} & -\frac{2 G^2 R^3 \text{$\rho $w}}{\left(1+G R^2\right)^2}+\frac{2 G R \text{$\rho $w}}{1+G R^2} & -D_\mathrm{W} k^2-\text{$\delta $w}
\end{array}
\right)
&=&\nonumber\\
\left(
\begin{array}{ccc}
 -5.01-1. k^2 & -1.36 & 9.80\times10^{-5} \\
 9.83 & 0.724 -0.008 k^2 & 0 \\
 7.08 & 1.96 & -0.5-1. k^2
\end{array}
\right)
\end{eqnarray}
 has at least one eigenvalue with positive real part, $\lambda=0.111$, with wavenumber $k=6$.
 
\subsection{Numerical Solution of PDEs}
The partial differential equations (PDEs) were solved with a finite element method as implemented in COMSOL Multiphysics 4.2a and 4.3. Details on how these models are implemented in COMSOL have been described by us recently \cite{Menshykau:2012vg,Germann:bT_kMV7D}. COMSOL Multiphysics is a well-established software package and several studies confirm that COMSOL provides accurate solutions to reaction-diffusion equations both on constant\cite{cutress2010} and growing domains\cite{Carin2006, Thummler2007, Weddemann2008}. We have previously used COMSOL to solve a similar mathematical set of PDEs on a bud shaped domain \cite{Menshykau:2012kg}. Table \ref{tbl:conv} shows that the converged solution differs by at most $5\%$ as the mesh is further refined. When simulations were performed on an open domain the COMSOL 'bulk solution condition' was implemented at a distance $6\sqrt{D\tau_\mathrm{max}}$ from the mesenchyme, where $\tau_\mathrm{max}$ is the maximum time of model evaluation. It has previously been shown that beyond this, the effects of diffusion is  not important on the experimental time scale \cite{Bardbook}. When simulating patterns on a growing domain, growth was prescribed to be normal to the boundary and proportional to the local level of signalling $\overline{v}_g=\overline{n}GR^2$, where $\overline{n}$ denotes unit vector normal to the surface \cite{Iber:2013vf}. 

\subsection{Robustness to Parameter Variability.} To estimate the robustness to spatial parameter variability we used a similar approach as in \cite{Bollenbach:2008p3894}. Thus, parameter values were assumed to be given by the formula $k=k_0\times(1+\xi_\mathrm{\theta}(x,y,z)))$, where $\xi_\mathrm{\theta}(x,y,z)$ is normally distributed random function with a mean value of zero and half width $\theta$. The half width of the distribution was equal for all parameters, except geometrical which were not varied.

\section*{Acknowledgements}
The authors are grateful to Alicia Kaestli for conducting some preliminary modelling, and to ETH Zurich for an ETH Fellowship to DM.

\section*{References}

\bibliography{papers}

\clearpage

\clearpage
\section*{Tables}
\begin{table}[h]
\begin{center}
\caption{\textbf{Values of Dimensionless Parameters used for the Simulations.} The domain geometry parameters were: $l_\mathrm{ep}=0.15$, $r_\mathrm{0}$= 0.5, $h_\mathrm{0}$=0.6. For those parameters marked by star (*) model read-outs were virtually independent of the exact value of these parameters. \label{tbl:paramD}}
\doublespacing
\begin{tabular}{| c  | c c | c |}
\hline
parameter  & epithelium & mesenchyme  & definition  \\

\hline
$D_\mathrm{W}$*     & 1  & 1 & WNT11 diffusion coefficient \\
$D_\mathrm{R}$     & 0.008  & --  &  RET diffusion coefficient\\
\hline
$\rho_\mathrm{G0}$ & -- & 0.18 &  GDNF production rate constant \\
$\rho_\mathrm{G}$ & -- & 2 & GDNF production rate constan\\
$\rho_\mathrm{W0}$* & 0.5 & -- & WNT11 production rate constan\\
$\rho_\mathrm{W}$ & 20 & -- & WNT11 production rate constan\\
$\rho_\mathrm{R}$ & 2 & -- & RET production rate constan\\
\hline
$\delta_\mathrm{C}$ & 0.5 & -- & complex formation rate constant \\
$v$	& 2.5 & -- &  RET production rate constant in response to RET signalling\\
\hline
$\delta_\mathrm{G}$* & 0.1 & 0.1 &  GDNF degradation rate constant\\
$\delta_\mathrm{W}$ & 6 & 6 & WNT11 degradation rate constant \\
$\delta_\mathrm{R}$ & 2 & 0 & RET degradation rate constant\\
\hline
\end{tabular}
\vskip10pt
\end{center}

\end{table}

\clearpage
\begin{table}
\centering
\caption{\textbf{Dimensional Parameter Values} To convert the dimensionless parameter values to their dimensional counterparts we used the known length and time scales of the process. Accordingly we set the characteristic length to $r_1 = 100~\mu m$ and the characteristic time to  $\frac{\overline{r}_1^2}{\overline{D}_G}=1000~s$. The pattern-forming mechanism is not sensitive to the exact protein concentrations as long as the relative ratios are preserved; since quantitative protein concentration measurements are not available for the ureteric bud we do not convert the parameter values that depend on protein concentrations. } \label{tbl:paramVal}
\doublespacing
\begin{tabular}{| c | c | c | c| c | c}
\hline
parameter  & value & experimental range & Definition  & References \\
\hline 
 $\overline{r}_0, \mu m$ & 50 & $\approx 50$ & outer diameter of the kidney mesenchyme (Fig. \ref{fig1} B)  & \cite{Muller:1997wl, Lu:2009p43731, Watanabe:2004kr, Chi:2009p43739}\\
$\overline{r}_1, \mu m$ & 100 &  $\approx 100$  & inner diameter of the kidney mesenchyme (Fig. \ref{fig1} B) & \cite{Muller:1997wl, Lu:2009p43731, Watanabe:2004kr, Chi:2009p43739}\\
$\overline{l}_\mathrm{ep}, \mu m$ & 10& $ \approx 15$ & epithelium thickness (Fig. \ref{fig1} B)  &  \cite{Muller:1997wl, Lu:2009p43731, Watanabe:2004kr, Chi:2009p43739}\\
\hline
$\overline{D}_\mathrm{G}, \mu m^2s^{-1}$ & 10 &   0.1-50 & GDNF diffusion coefficient   & \cite{Kicheva2007, Yu2009} \\
$\overline{D}_\mathrm{W}, \mu m^2s^{-1}$ &  10 & 0.1-50 & WNT11 diffusion coefficient  & \cite{Kicheva2007, Yu2009} \\
$\overline{D}_\mathrm{R}, \mu m^2s^{-1}$ &  0.08 & 0.001-0.5 & RET diffusion coefficient & \cite{Kumar2010, Hebert2005, Ries2009}\\
$\overline{D}_\mathrm{ext}^d, \mu m^2s^{-1}$ &  100  & 10-200 & diffusion constant outside the tissue & \cite{Nesmelova2002} \\
\hline
$\overline{\delta}_{\mathrm{G}}, s^{-1}$ &  $1\times10^{-4}$ & $10^{-4}-10^{-3}$ & GDNF degradation rate constant &  \cite{Yu2009, Kicheva2007} \\
$\overline{\delta}_{\mathrm{W}}, s^{-1}$ & $6\times10^{-3}$ & $10^{-4}-10^{-3}$&  WNT11 degradation rate constant &\cite{Yu2009, Kicheva2007} \\
$\overline{\delta}_{\mathrm{R}}, s^{-1}$ & $2\times10^{-3}$ & $10^{-4}-10^{-3}$& RET degradation rate constant &\cite{Yu2009, Kicheva2007} \\
\hline
\end{tabular}\\[0.25cm]
\end{table}

\clearpage
\section*{Tables}
\begin{table}[h]
\begin{center}
\caption{\textbf{Accuracy of the Numerical Solution.} Maximal values of $R^2G$ at the epithelium-mesenchyme border and corresponding size of the numerical problem (here $k$ denotes thousands degrees of freedom), calculated  for various mesh sizes in the epithelium and in the mesenchyme layer. The value marked in red indicates the mesh used to calculate the data in this manuscript.} \label{tbl:conv}
\doublespacing
\begin{tabular}{ c  | c | c | c | c | c | c }
\hline
&  & \multicolumn{5}{c}{FEM mesh size, epithelium}    \\
\hline
	& & 0.2  & 0.1 & 0.06 & 0.04 & 0.03  \\
\hline
\multirow{4}{*}{FEM mesh size, mesenchyme}  & 0.3 &	58.25/3.1k	&49.69/11.7k	&46.05/35.3k	&45.08/99.2k&	44.60/211k \\
&0.2	&53.04/4.06k&	50.08/11.8k&	\textcolor{red}{46.63/35.2k}&	45.1/99.4k&	44.53/211k\\
&0.1	&49.98/25.4k	&48.37/25.7k	&46/46.2k	&45.01/106k&	44.61/217k\\
&0.06&	46.03/105k&	45.35/105k	&45.94/113k	&45.04/167k&	44.41/272k\\
\hline
\end{tabular}
\vskip10pt
\end{center}
\end{table}

\clearpage

\clearpage

\clearpage

\clearpage

\clearpage

\newpage

\end{document}